\documentclass{llncs}

\usepackage[textwidth=15cm,textheight=24cm,centering]{geometry}

\usepackage{graphicx}
\usepackage{amssymb,amsmath,graphicx,algorithm2e}

\newcommand{\st}{\mbox{subject to}}

\begin{document}



\title{iTreePack: Protein Complex Side-Chain Packing by Dual Decomposition}


\author{Jian Peng\,$^{1}$, Raghavendra Hosur\, $^{2}$, Bonnie Berger\, $^{2*}$ and Jinbo Xu $^{3*}$}

\institute{Universit of Illinois at Urbana-Champaign \and Toyota Technological Institute at Chicago\and
Computer Science and Artificial Intelligence Laboratory, MIT\\
Correspondence to bab@csail.mit.edu or j3xu@ttic.edu}

\maketitle

\begin{abstract}
Protein side-chain packing is a critical component in obtaining the 3D coordinates of a structure and drug discovery. Single-domain protein side-chain packing has been thoroughly studied. For instance, our efficient tree decomposition algorithm TreePack has been re-implemented in SCWRL, the widely-used side-chain packing program, for monomer side-chain packing. A major challenge in generalizing these methods to protein complexes is that they, unlike monomers, often have very large treewidth, and thus algorithms such as TreePack cannot be directly applied. To address this issue, SCWRL4 treats the complex effectively as a monomer, heuristically excluding weak interactions to decrease treewidth; as a result, SCWRL4 generates poor packings on protein interfaces. To date, few side-chain packing methods exist that are specifically designed for protein complexes.\\
In this paper, we introduce a method, iTreePack, which solves the side-chain packing problem for complexes by using a novel combination of dual decomposition and tree decomposition. In particular, iTreePack overcomes the problem of large treewidth by decomposing a protein complex into smaller subgraphs and novelly reformulating the complex side-chain packing problem as a dual relaxation problem; this allows us to solve the side-chain packing of each small subgraph separately using tree-decomposition. A projected subgradient algorithm is applied to enforcing the consistency among the side-chain packings of all the small subgraphs. Computational results demonstrate that our iTreePack program outperforms SCWRL4 on protein complexes. In particular, iTreePack places side-chain atoms much more accurately on very large complexes, which constitute a significant portion of protein-protein interactions. Moreover, the advantage of iTreePack over SCWRL4 increases with respect to the treewidth of a complex. Even for monomeric proteins, iTreePack is much more efficient than SCWRL and slightly more accurate.

\end{abstract}


\section{Introduction}


PPIs (protein-protein interactions) or protein complexes specify physical interactions between pairs or groups of proteins. Many
fundamental cellular processes such as DNA replication, transcription, translation and signal transduction
are mediated through a complex network of PPIs \cite{PPIdata,Interactome}. High-throughput experimental methods have been used
to determine PPI networks for model organisms including yeast\cite{PPIyeast1,PPIyeast2} and human\cite{PPIhuman}. For example, ~30k PPIs are discovered for ~6200 yeast proteins. Computational methods are also developed for PPI predictions \cite{Struct2Net,iWRAP,LTHREADER,PPIPotential,MULTIPROSPECTOR,BENHUR,NOBLE,WANG_PPI,Valencia02a,Nimwegen,Jiang}.
These methods usually can only identify whether two proteins interact or the composition of a
PPI, but not the atomic structures or binding modes of PPIs. The 3D structures of PPIs are important
for the understanding of cellular processes at molecular level. Atomic structures of PPIs are also
important for developing drugs targeting PPIs. Recently, many efforts have also been dedicated to
protein interface design \cite{DrugDesign1,DrugDesign2}.
The atomic detail of the interaction between proteins usually determines the affinity and specificity for binding.
In order to obtain atomic structure of a PPI, side-chain packing is an indispensable step.
Side-chain packing places side-chain atoms assuming the backbone structure is given.
Although there are many algorithms for protein side-chain packing \cite{TreePack,SCWRL4,SCWRL3,DEE,ILP1,ILP2,DockingSideChain,BakerSC,Honig,OPUS,RASP,LiangS,SDP},
few of them has been focused on protein complexes \cite{DockingSideChain,BakerSC}, especially the interfacial regions.

Side-chain packing problem can be formulated as a combinatorial optimization problem.
Many computational techniques have been studied for this problem,
such as integer linear programming \cite{ILP1,ILP2}, dead-end-elimination \cite{DEE} and graph decomposition \cite{RECOMBTreePack,TreePack,SCWRL3}.
Xu first introduced tree-decomposition algorithm \cite{RECOMBTreePack} with a non-trivial time complexity
that can find the globally optimal solution.
This algorithm has been reimplemented by Dunbrack as the major energy optimizer of SCWRL4,
possibly the most widely-used side-chain packing program \cite{SCWRL4}.
Nevertheless, the computational and space complexity of the tree-decomposition algorithm are exponential with respect to the treewidth of the underlying residue interaction graph of a protein.
Treewidth is a parameter measuring the topological complexity of the graph \cite{GraphMinor,GraphMinor2} for each protein.
Single-domain or monomer proteins usually have small treewidth (less than 20) and thus, are amenable to tree-decomposition.
However, the tree-decomposition algorithm may not be directly applied to complex side-chain packing
since complexes usually have large treewidth.
To deal with proteins of large treewidth, SCWRL4 heuristically excludes
some weak but important interactions from consideration in order to have an interaction graph with small treewidth.
However, such a heuristic sometimes fails to place side-chain atoms accurately due to ignorance of important interactions
especially when the protein complexes under consideration have very large treewidths. To date, efficient and accurate side-chain packing methods, specifically designed for protein complexes, do not exist.


In this paper, we present iTreePack, an efficient dual decomposition algorithm for protein complex side-chain packing.
Instead of directly applying tree-decomposition to a complex with large treewidth, iTreePack solves the dual relaxation of the problem through a novel combination of dual decomposition and tree decomposition.
The dual problem is constructed using a vertex duplication technique,
which divides a large interaction graph into smaller interaction subgraphs of much smaller treewidth,
each corresponding to one monomer or protein interface.
iTreePack solves side-chain packing for each subgraph separately and efficiently using tree-decomposition since the treewidth is small.
To ensure the consistency among side-chain packings of the subgraphs,
we use a projected subgradient algorithm to update the side-chain packing of each subgraph iteratively.
The consistency usually can be achieved within 10-20 iterations.
Therefore, this algorithm can accurately place side-chain atoms for complexes with very large treewidth
without excluding important interactions from consideration.

Computational results demonstrate that our iTreePack program noticeably outperforms the widely-used SCWRL4 when applied to complexes, considering most algorithmic advances in this field only lead to marginal improvements. In particular, iTreePack places side-chain atoms much more accurately on very large complexes (treewidth greater than 20), which constitute a significant portion of protein-protein interactions. Moreover, the advantage of iTreePack over SCWRL4 increases with respect to the treewidth of a complex, especially on the protein interfaces. Even for monomeric proteins, iTreePack is much more efficient than SCWRL and slightly more accurate.

{\footnotesize
\section{Notations and Problem Setting}
For notational simplicity, we assume that a protein complex consists of only a pair of interacting monomers.
However, our algorithms and results also readily apply to a complex of multimers.

\subsection{Residue Interaction Graph (RIG) for Monomers}
For each monomer structure $A$, we use a residue interaction graph $G_A=(V_A,E_A)$
to represent the residues in a protein and their interaction relationship.
Each vertex in the graph represents a residue associated with the 3D coordinates of the corresponding residue center.
Let $D_A[i]$ denote the set of all possible rotamers for residue $i$.
There is an edge $(i,j)\in E_A$ between two residues $i$ and $j$ if and only if there are two rotamers $l\in D_A[i]$ and $k\in D_A[j]$
such that at least one atom in rotamer $l$ is in contact with at least one atom in rotamer $k$.
Two atoms are in contact if and only if their Euclidean distance is less than a constant $D_u$.

For each rotamer $k\in D_A[i]$, we use a singleton score $S_{A_i}(k)$
to describe the preference of assigning the rotamer $k$ to the specific residue $i$.
For any two rotamers $l\in D_A[i]$ and $k\in D_A[j]$, we assign a score $P_{A_i,A_j}(l,k)$ to describe their pairwise interaction energy.
When there is no edge between two residues $i$ and $j$, $P_{A_i,A_j}(l,k)$ is equal to 0 for any $l$ and $k$.
The more detailed description of singleton and pairwise scores will be presented in section \ref{subsec:energy}.

\subsection{Protein Interface Graph (PIG) for Protein Interfaces}
%

To model interactions between monomers $A$ and $B$, we introduce a protein interface graph $G=(V_{IA},V_{IB},E_{AB})$
where $V_{IA}$ and $V_{IB}$ represents all interfacial residues in monomers $A$ and $B$, respectively,
and $E_{AB}$ denotes the set of interfacial edges between $A$ and $B$.
. A residue is an interfacial residue if its distance ($C_{\beta}$ atoms are used to calculate distance) to the closest residue in the other monomer is less than a constant $D_{int}$. Each vertex $i\in V_{IA}$ (or $j\in V_{IB}$) represents an interfacial residue in monomer $A$ (or $B$). There is an edge $(i,j)\in E_{AB}$ between two residues $i \in V_{IA}$ and $j\in V_{IB}$ if and only if their $C_{\beta}$ atoms are within distance $D_{int}$.
Therefore, a protein interface graph is bipartite since the two ends of one edge must be in different monomers.
Similar to residue interaction graph, we also have pairwise score $P_{A_i,B_j}(l,k)$ for each rotamer pair of two interfacial residues.

\subsection{Rotamer Library and Energy Function}\label{subsec:energy}

{\bf Rotamer Library.} Since the local structures of protein interfaces are usually more flexible than the core regions of globular proteins, we use slightly different libraries for interfacial residues and non-interfacial residues.
For non-interfacial residues, we use the popular backbone-dependent rotamer library \cite{RotamerLibrary} developed by Dunbrack. For each residue, we  calculate the backbone $\phi$ and $\psi$ angles and retrieve the backbone-dependent rotamer library from \cite{RotamerLibrary}.
Each rotamer is associated with its occurring frequency $f_{A,i}(k)$, which is also taken from Dunbrack's rotamer library.

For interfacial residues, we augment the rotamer library by considering a larger range of backbone angles.
For each residue, in addition to the rotamers associated with backbone $\phi$ and $\psi$,
we also use auxiliary rotamers associated with $\phi\pm 10^{\circ}$ and $\psi\pm 10^{\circ}$ to increase conformation space at the interface region.
The frequency of each auxiliary rotamer is reduced by half of its frequency in Dunbrack's rotamer library to avoid introducing too much noise.

\noindent {\bf Statistical Local Potential.}
We use similar statistical local potentials for all rotamers \cite{SCWRL3}.
\vspace{-0.2cm}\begin{equation*}
Local_{A_i}(l)=-K\log \frac{f_{A,i}(l)}{\max_{k\in i}f_{A,i}(k)}
\end{equation*}
K is optimized to 10 to yield the best prediction accuracy.

\noindent{\bf CHARMM non-bonded potential.}
The energy functions used in original TreePack and SCWRL are approximations of the Lennard-Jones potential for van der Waals component.
In this work, we directly use the Lennard-Jones potential $LJ(a,b)$ and also the electrostatic interaction for each atom pair $a$ and $b$.
All the parameters of the non-bonded potentials are from the CHARMM19 force field.

\noindent {\bf Singleton Score.}
Let $bb_A(i)$ be the set of backbone atoms of for residue $i\in V_A$; $sc_{A_i}(l)$ be the set of all side-chain atoms of rotamer $l \in D_A[i]$; $dist(a,b)$ be the Euclidean distance between two atoms $a$ and $b$; $NonBond(a,b)=LJ(a,b)+Elec(a,b)$ as the CHARMM non-bonded potential between atoms $a$ and $b$. The singleton score for each rotamer $l\in D[i]$ is calculated as follows.
\begin{equation*}
S_{A_i}(l) = Local_{A_i}(l) + \sum_{u\in sc_{A_i}(l)}\sum_{j\in V_A}\sum_{a\in bb_A(j)} NonBond(u,a)
\end{equation*}
For the interfacial residues, the singleton score also includes the non-bond potential with backbone atoms in the other monomer.
\noindent {\bf Pairwise Score.}
For any pair of interacting residues $i$ and $j$ and their two rotamers $k\in D_A[i]$ and $l\in D_B[j]$, the pairwise score is the sum of non-bonded potentials of all side-chain atom pairs in the two rotamers.
\begin{equation*}
P_{A_i,A_j}(k,l) = \sum_{u\in sc_{A_i}(l)}\sum_{v\in sc_{A_i}(k)} NonBond(u,v)
\end{equation*}
The pairwise scores between two rotamers of different monomers are computed similarly.

\subsection{Mathematical formulation}
The complex side-chain packing problem can then be formulated as a combinatorial optimization problem on a large complex residue interaction graph.
Let $R=(R_A, R_B)$ denote the side-chain packing of a complex where $R_A$ and $R_B$ are the packings for the constituent monomers $A$ and $B$, respectively. Meanwhile, $R_A(i)$ and $R_B(j)$ denote the rotamers chosen for residue $i$ in monomer $A$ and residue $j$ in monomer $B$, respectively.
The optimal side-chain packing minimizes the following energy function.
\begin{equation}
G(R_A,R_B) = G_A(R_A) + G_A(R_B) + G_{AB}(R_{A},R_{B})
\label{eq:ori}
\end{equation}
where $G_A(R_A)= \sum_{i\in V_A} S_{A_i}(R_A(i)) + \sum_{(i,j)\in E_A} P_{A_i,A_j}(R_A(i),R_A(j))$ and $G_B(R_B)=\sum_{i\in V_B} S_{B_i}(R_B(i)) + \sum_{(i,j)\in E_B} P_{B_i,B_j}(R_B(i),R_B(j))$ are the intra-monomer energy functions and $G_{AB}(R_{A},R_{B}) = \sum_{(i,j)\in E_{AB}} P_{A_i,B_j}(R_A(i),R_B(j))$ is the interfacial energy function.

\section{Methods}
\subsection{Tree Decomposition for Monomer Side-Chain Packing}
The notions of tree-decomposition and treewidth were introduced by Robertson and Seymour in their seminal work on graph minor theory\cite{GraphMinor,GraphMinor2}, which captures the structural features of all graphs excluding a given minor.
The complexity of a graph can be measured by its treewidth, which is the minimum width over all possible tree-decompositions.
The width of a tree-decomposition is the maximum component size minus 1.
Many optimization problems can be solved using graph tree-decomposition with a time complexity polynomial in graph size when the treewidth is small. Tree-decomposition has been successfully applied to monomer side-chain packing \cite{TreePack},
resulting in the first subexponential-time algorithm, which is linear when treewidth is small, for this problem.
Afterwards, tree-decomposition has been applied to many biology problems including contact map overlap\cite{ContactMap},
protein alignment\cite{StrAln}, RNA analysis\cite{Sequencing}, de novo sequencing \cite{Sequencing} and network analysis \cite{NetAnalysis}.

Although the problem of finding an optimal tree decomposition of a general graph is NP-hard,
a variety of heuristic algorithms can produce near optimal treewidth in practice, e.g., the minimum fill-in heuristic algorithm.
Given a tree decomposition of a protein/complex interaction graph (RIG or PIG),
dynamic programming can be applied to computing the optimal side-chain packing of the protein in two major steps: bottom-to-top and top-to-bottom.
The bottom-to-top step is used to calculate the optimal energy function and the top-to-bottom step is used to extract the optimal
side-chain packing. This is analogous to dynamic programming for sequence alignment where a forward step is used to calculate
the optimal alignment score and the backward step is used to extract the optimal alignment.
More detailed account of the tree decomposition algorithm for monomer side-chain packing is described in \cite{TreePack}.

The computational complexity of the tree decomposition algorithm for side-chain packing is first analyzed in \cite{RECOMBTreePack}.
The complexity of the algorithm is mainly determined by the treewidth of the graph, which is bounded by $O(|V|^{\frac{2}{3}}\log |V|)$ from above, where $|V|$ is the number of vertices in the graph. The main result is summarized in the following theorem \cite{TreePack}.
\begin{theorem}
\label{th:tw}
The tree decomposition based side-chain packing algorithm has time complexity $O(Nn^{tw+1}_{rot})$ where $N$ is the length of the protein, $n_{rot}$ the average number of rotamers for each residue, and $tw$ the treewidth of the residue interaction graph which is bounded by $O(N^{\frac{2}{3}}\log N)$.
\end{theorem}

The tree-decomposition algorithm works well for monomers, which usually has small treewidth.
However, a protein complex has a much larger treewidth.
Theoretically, a protein complex  of $m$ monomers (each with $N$ residues) could have treewidth $O(m^{\frac{2}{3}}N^{\frac{2}{3}}(\log m + \log N))$.
 Since the time complexity of the tree-decomposition algorithm is exponential with respect to treewidth,
the tree-decomposition algorithm becomes very expensive or infeasible for a complex with very large treewidth.
That is, a much better algorithm is needed to address the complex side-chain packing problem.

\subsection{Dual Decomposition for Complex Side-Chain Packing}
Here we present technical details of iTreePack, an efficient optimization algorithm
for complex side-chain packing by dual decomposition.
Dual decomposition has been applied to many computationally expensive problems,
such as natural language processing \cite{DualNLP} and inference of graphical models \cite{DualSurvey}.
The challenge of complex side-chain packing problem by tree-decomposition lies in large treewidth of a complex.
However, the treewidth of a monomer or a protein interface usually is quite small.
The key idea of dual decomposition is to decompose the large complex graph into several subgraphs with small treewidth and then
apply the tree-decomposition algorithm to each subgraph separately.
To achieve consistency among the solutions of the subgraphs,
we iteratively adjust the weight of each subgraph according to the degree of consistency among solutions using a subgradient descent algorithm.
At each iteration, we can also estimate the gap between current solution and the optimal solution.
Finally we can reach an optimal or near-optimal solution of the complex side-chain packing problem.

\noindent {\bf Complex Decomposition through Vertex Duplication.}
To decompose a protein complex graph into subgraphs,
we make two copies of each interfacial residue, one belonging to a monomer and the other to a protein interface.
See Figure \ref{fg:vertex_dup} for an example.
To avoid overcounting the singleton score of interfacial residues, we assign the full score
to the monomeric copy and 0 to the interfacial copy.
By this way, we can separate a large complex residue interaction graph to
a few monomer residue interaction graph plus some protein interface graphs.
Each subgraph potentially has much smaller treewidth than the original complex graph.

\begin{figure}
\begin{center}
\begin{tabular}{cc}
\includegraphics[width=0.4\columnwidth]{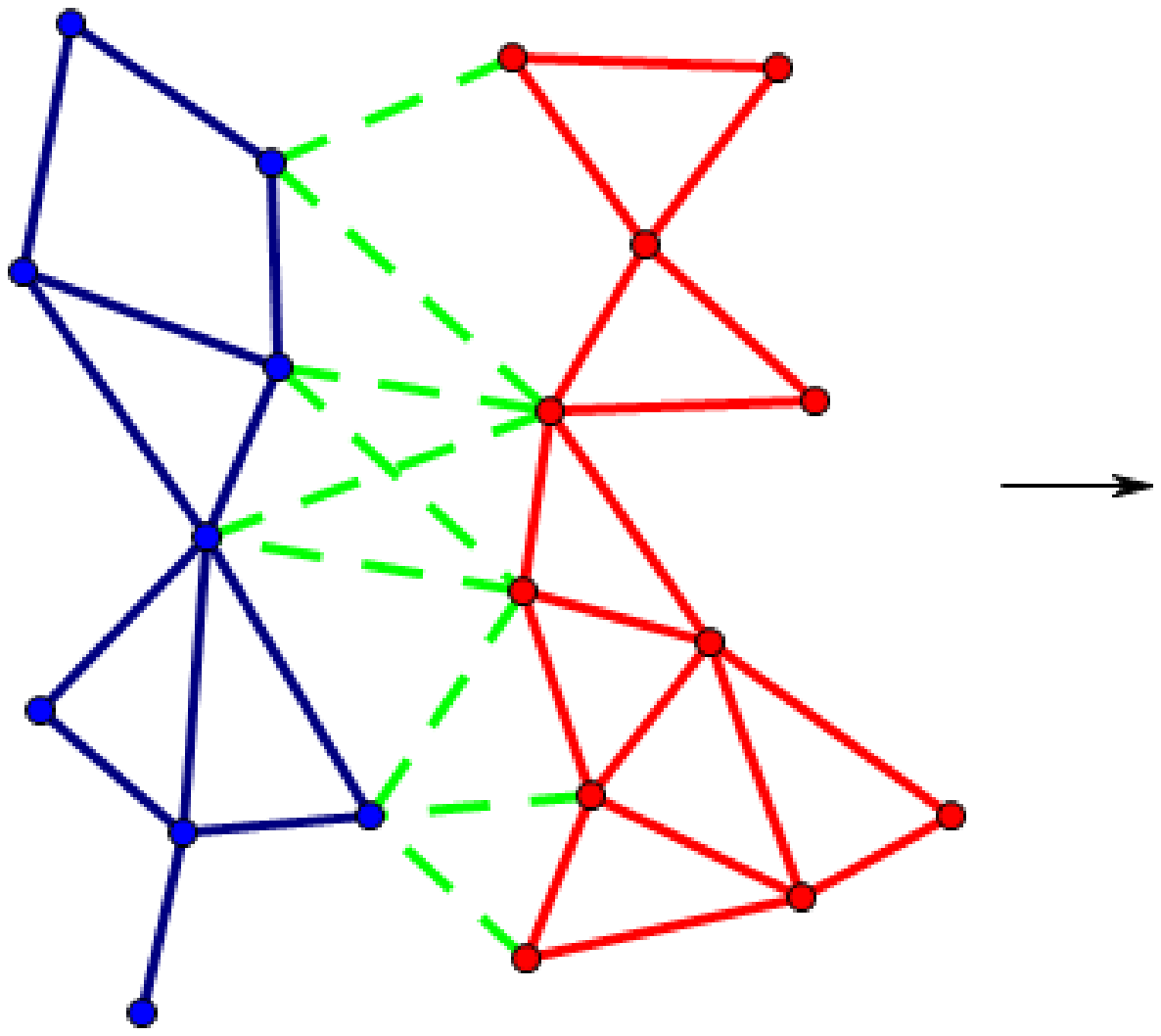} &
\includegraphics[width=0.5\columnwidth]{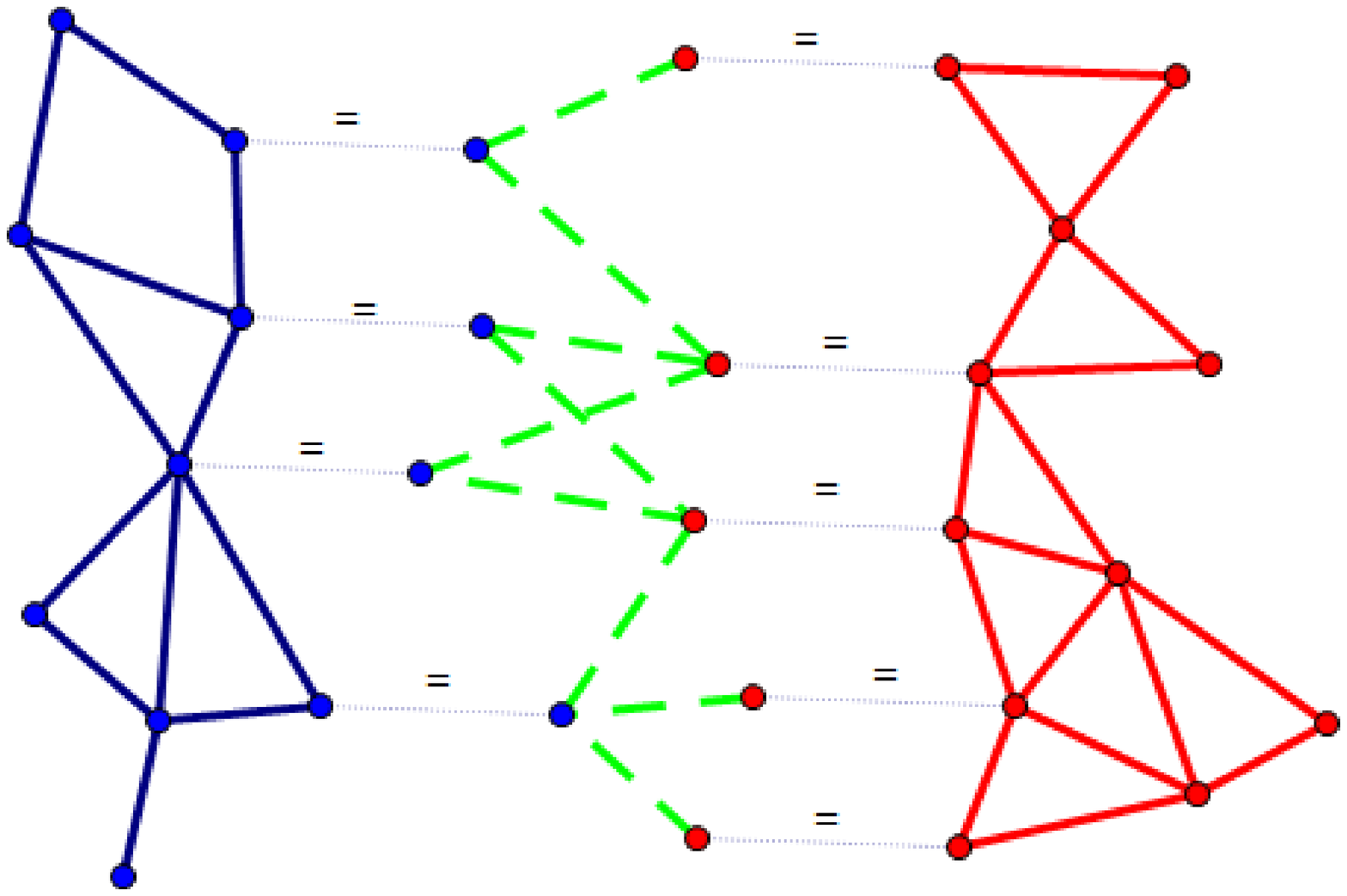} \\
\end{tabular}
\caption{An example of graph decomposition of a dimeric protein complex.
Each interfacial residue is duplicated once.
The two monomer residue interaction graphs are in green and red and the protein interface graph is in blue.}
\label{fg:vertex_dup}
\end{center}\vspace{-1cm}
\end{figure}
\noindent {\bf Dual Relaxation and Decomposition.}
Let $R_{IA}$ and $R_{IB}$ denote rotamer assignments for interfacial residues in $A$ and in $B$, respectively.
The problem formulation in Eq. (\ref{eq:ori}) can be rewritten as follows.
\begin{equation}
\label{eq:opt}
\begin{array}{cl}
\mathop{\min\limits_{R_A,R_B,R_{IA},R_{IB}}} & G_A(R_A) + G_B(R_B) + G_{AB}(R_{IA},R_{IB})\\
\st & R_{IA}(j) = R_A(j) \quad \mbox{for all} \quad j \in V_{IA}\\
    & R_{IB}(k) = R_B(k) \quad \mbox{for all} \quad k \in V_{IB}\\
\end{array}
\end{equation}
It is not hard to prove that the formulation in Eq. (\ref{eq:opt}) is equivalent to the formulation in Eq.(\ref{eq:ori}).

Since the optimization problem (\ref{eq:opt}) is still intractable and as hard as problem (\ref{eq:ori}), we do not solve it directly.
Instead we construct a Lagrangian dual problem by relaxing the equality constraints in (\ref{eq:opt}).
In particular, we introduce a Lagrangian variable for each equality constraint in (\ref{eq:opt}) to
obtain the following dual problem.
\vspace{-0.1cm}\begin{equation}
\label{eq:dual}
\begin{array}{rl}
\hspace{-1.5cm}& L(u)= L(u_{IA},u_{IB}) = \\
&\mathop{\min\limits_{R_A,R_B,R_{IA},R_{IB}}} \{G_A(R_A) + G_B(R_B) + G_{AB}(R_{IA},R_{IB}) \\
& + \mathop{\sum\limits_{j\in V_{IA}}}\mathop{\sum\limits_{l\in D_A[j]}} u_{IA}[j](l)(\delta(R_{A}(j)=l) - \delta(R_{IA}(j)=l)) \\
& + \mathop{\sum\limits_{j\in V_{IB}}}\mathop{\sum\limits_{l\in D_B[j]}} u_{IB}[j](l)(\delta(R_{B}(j)=l) - \delta(R_{IB}(j)=l)) \} \\
&= \quad \mathop{\min\limits_{R_A}} \{ G_A(R_A) + F_A(u_{IA},R_A) \} + \mathop{\min\limits_{R_B}} \{ G_B(R_B) + F_B(u_{IB},R_B) \}\\
&  + \mathop{\min\limits_{R_{IA},R_{IB}}} \{G_{AB}(R_{IA},R_{IB}) - H(u_{IA},R_{IA},u_{IB},R_{IB}) \}\vspace{-0.3cm}
\end{array}
\end{equation}
where \vspace{-0.3cm}\begin{enumerate}
\item $u_{IA}[j](l)$ is the dual variable for the equality constraint over rotamer $l\in D_A[j], j\in V_{IA}$.
\item $\delta(R_{A}(j)=l)$ is an indicator function, equal to 1 if $R_A(j)=l$; otherwise 0.
\item $F_A(u_{IA},R_A)=\mathop{\sum\limits_{j\in V_{IA}}}\mathop\sum\limits_{l\in D_A[j]} u_{IA}[j](l)(\delta(R_{A}(j)=l))$
and $F_B(u_{IB},R_{IB})$ is defined similarly.
\item
$H(u_{IA},R_{IA}, u_{IB}, R_{IB})={\sum_{j\in V_{IA}}}{\sum_{l\in D_A[j]}} u_{IA}[j](l) \\ \delta(R_{IA}(j)=l)+
\sum_{j\in V_{IB}}{\sum_{l\in D_B[j]}} u_{IB}[j](l)\delta(R_{IB}(j)=l)$.
\end{enumerate}

It is not difficult to prove that problem (\ref{eq:dual}) is a relaxation of the problem (\ref{eq:opt}).
That is, $L(u_{IA},u_{IB})$, also denoted as $L(u)$, is a lower bound of the objective function of the problem (\ref{eq:opt}).
Problem (\ref{eq:dual}) consists of three independent and small subproblems,
each of which corresponds to a subgraph generated by complex decomposition
and can be solved efficiently and separately by tree-decomposition.
Therefore, we can solve problem (\ref{eq:dual}) and then use its solution to approximate the problem (\ref{eq:opt}).
Since the equality constraints in problem (\ref{eq:opt}) are relaxed,
the side-chain packings of the subproblems may be inconsistent in the interface regions.
That is, the solutions $R_A$, $R_B$, $R_{IA}$, $R_{IB}$ may not satisfy the equality constraints in problem (\ref{eq:opt}).
See \cite{DualSurvey} for more detailed account of theoretical analysis and optimality conditions of the dual relaxation approach.

To make the side-chain packings of all the subproblems consistent with one another
and also to approach to the optimal side-chain packing solution,
we maximize $L(u)$ with respect to $u$.
Since $L(u)$ is a piecewise linear function over dual variable $u$,
it can be optimized by the projected subgradient algorithm, which improves $L(u)$ iteratively.
The subgradient of the dual variable can be calculated directly from the solutions of subproblems.
At each iteration, the subproblems are updated by the subgradient of dual variables and re-optimized to obtain new subgradient.
Finally, $L(u)$ will be equal to or very close to the optimal solution of problem (\ref{eq:opt})
and the side-chain packings of all the subgraphs will be (almost) consistent with one another.
The overall algorithm is shown in Algorithm \ref{alg:dualdecomp}.

At each iteration of the projected subgradient algorithm,
we can construct a feasible solution to the primal problem (\ref{eq:opt}).
That is, we can obtain such a feasible solution by assembling the side-chain packings of all the constituent monomers
and ignoring the packings of all the protein interfaces.
Such a feasible solution bounds the optimal solution of problem (\ref{eq:opt}) from above.
Our dual decomposition can terminate when the gap between $L(u)$ and this feasible solution is sufficiently small
or the algorithm itself is converged.
Our experiments indicate that it takes up to 20 iterations to terminate the projected subgradient algorithm.
\begin{algorithm}\vspace{-0.2cm}
\rule{3in}{0.01cm}\\
{\bf Initialization:} $u_{IA}=0$, $u_{IB}=0$ for all rotamers of interfacial residues\\
\vspace{0.1cm}
\While{not converged}{
\vspace{0.1cm}
1. Solve $R_A^*=\arg\min_{R_A}{G_A(R_A) + F_A(u_{IA},R_A)}$ by tree-decomposition\;
\vspace{0.1cm}
2. Solve $R_B^*=\arg\min_{R_B}{G_B(R_B) + F_B(u_{IB},R_B)}$ by tree-decomposition\;
\vspace{0.1cm}
3. Solve {$R_{IA}^*,R_{IB}^* = \arg\min_{R_{IA},R_{IB}} {G_{AB}(R_{IA},R_{IB}) - H(u_{IA},R_{IA}, u_{IB},R_{IB})}$}
by tree-decomposition\;
4. Update $\alpha_t$\;
5. For all $j\in V_{IA}$, if $R_A^*(j)\neq R_{IA}^*(j)$, $u_{IA}[i](R_A^*(j)) \leftarrow u_{IA}[i](R_A^*(j)) + \alpha_t(+1)$, \\
  \hspace{0.3cm} $u_{IA}[i](R_{IA}^*(j)) \leftarrow u_{IA}[i](R_{IA}^*(j)) + \alpha_t(-1)$.\\ \hspace{0.35cm}Update $u_{IB}$ in the same way\;
}
{\bf Output:} $R_A^*$ and $R_B^*$\\
\caption{Optimization Algorithm for the Dual Relaxation}
\label{alg:dualdecomp}
\rule{3in}{0.01cm}\vspace{0.3cm}
\end{algorithm}

\hspace{-0.5cm}{\bf Computational Complexity of Dual Decomposition.}
By Theorem \ref{th:tw},
the computational complexity of tree-decomposition for each subproblem is exponential with respect to the treewidth of each subgraph.
Usually the protein interface graph has a smaller treewidth,
the computational complexity is dominated by the tree decomposition algorithm on the monomer with the largest treewidth.
Supposing the maximum monomer has $N$ residues and a complex has $m$ monomers,
the computational complexity of the dual decomposition algorithm is $O(mNn^{O(N^{\frac{2}{3}}\log N)}_{rot})$
instead of $O(mNn^{O(m^{\frac{2}{3}}N^{\frac{2}{3}}(\log m + \log N))}_{rot})$.
Since $L(u)$ is piecewise linear and Lipschitz-continuous,
the projected subgradient algorithm converges within $O(\frac{1}{\epsilon^2})$ iterations if appropriate stepsize scheme is chosen,
where $\epsilon>0$ is the tolerance.
In summary, the computational complexity of the overall algorithm is $O(\frac{1}{\epsilon^2}mNn^{O(N^{\frac{2}{3}}\log N)}_{rot})$.

\subsection{Implementation Detail}
{\bf Dead End Elimination.}
To reduce the combinatorial conformation space, we use dead-end-elimination (DEE)
to remove rotamers that cannot lead to the optimal conformations.
We first employ the Goldstein criterion DEE \cite{Goldstein} and then split DEE \cite{SplitDEE} until no rotamers can be eliminated.
The DEE techniques are powerful and often reduce the conformation space substantially.

\noindent {\bf Stepsize update.}
Theoretically, the $O(\frac{1}{\epsilon^2})$ convergence rate can be guaranteed if the stepsize $\alpha_t$ of the projected subgradient method satisfies 1) $\lim_{t\rightarrow \infty}\alpha_t = 0$ and 2)$\sum_t \alpha_t = \infty$.
Empirically, we set $\alpha_t$ to $\mu{d_t}/(d_0)$
where $d_t$ is the number of violated equality constraints after iteration $t$ and $\mu$ is the initial stepsize.
This step size works very well and the dual decomposition algorithm terminates within 20 steps.
}

\section{Experimental Results}

\subsection{Protein Complex Benchmark}
To evaluate the performance of our method, we compiled a set of 547 protein complexes randomly selected from 3Dcomplex \cite{3DComplex}.
Each complex contains at least 10 interfacial residues and any two complexes share less than 30\% sequence identity.
These complexes contain 202 to 3084 residues.
In this benchmark, the maximal number of constituent monomers in a complex is 20 and the maximal number of interfaces is 16.
In the following sections, we use TreePack to denote the tree-decomposition algorithm in \cite{TreePack}
and iTreePack the work presented in this paper.
Note that TreePack uses the energy function described in this paper, which is different from the original TreePack \cite{TreePack}.
That is, both TreePack and iTreePack use the same energy function, which allows the fair comparison of their algorithms.

\subsection{Treewidth Distribution}
\begin{figure}
\hspace{-0.6cm}\includegraphics[width=1.2\columnwidth]{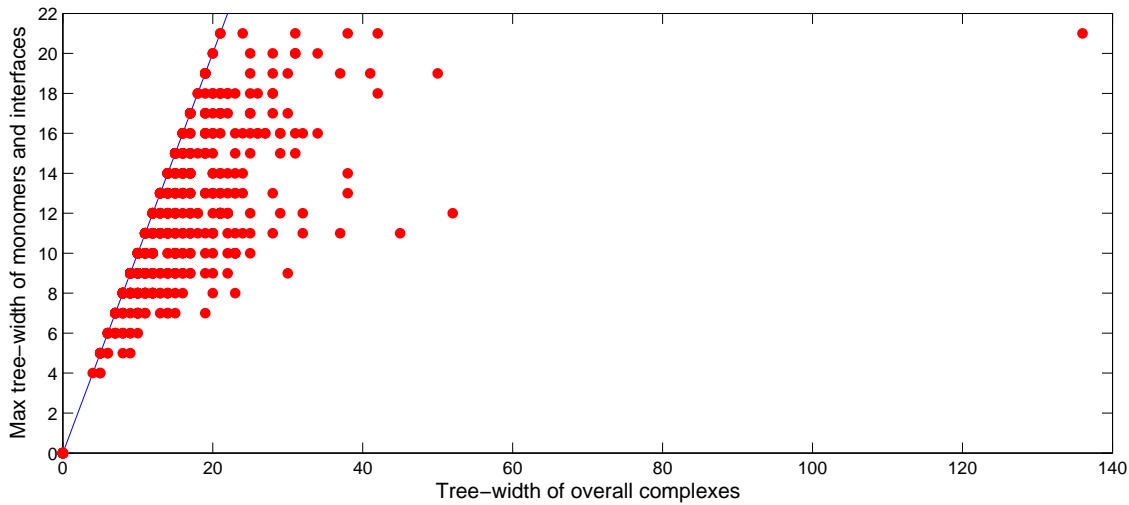}
\caption{Treewidth distribution of complex graphs and their subgraphs.}
\label{fg:TWS}\vspace{-0.5cm}
\end{figure}

We calculate the approximate treewidth of complex interaction graphs and monomer/interface interaction subgraphs using the minimum
fill-in heuristic algorithm \cite{MinFill}.
The interaction graphs is built after the dead-end-elimination steps with both $D_u$ and $D_{int}$ being set to 8$\dot{A}$.
Figure \ref{fg:TWS} shows the treewidth distribution of complex interaction graphs and the constituent subgraphs.
For each complex, only the maximum treewidth of its constituent monomers is shown.
In total, 293 of the 547 complexes contain subgraphs with smaller treewidth and
90 of them have treewidth at least 8 less than that of the complex itself.
The average treewidth of the complex interaction graphs is 15.26 while that of the largest subgraph treewidths is 11.26.
The largest protein complex $1jnb$, which consists of 3084 residues,
has treewidth 136 while its 12 constituent monomers have treewidths ranging from 17 to 21.
This observation indicates that by applying tree-decomposition to individual monomer separately,
we can significantly improve computational efficiently and save memory usage.

\subsection{Performance Evaluation}
We tested iTreePack and SCWRL4 \cite{SCWRL4} on all the 547 protein complexes.
The side-chain prediction for a residue is treated as correct if the deviation of
the predicted $\chi_1$ angle from the native is less than $40^{\circ}$.
As a control, we also ran TreePack and SCWRL4 on all constituent monomers separately without considering their interactions.
We use SCWRL4-monomer to denote the results obtained by running SCWRL4 on monomers.
By comparing SCWRL4-monomer with SCWRL4 and iTreePack, we can estimate the importance of protein
interfaces to complex side-chain packing.

\begin{figure}
\begin{tabular}{cc}
\includegraphics[width=0.50\columnwidth]{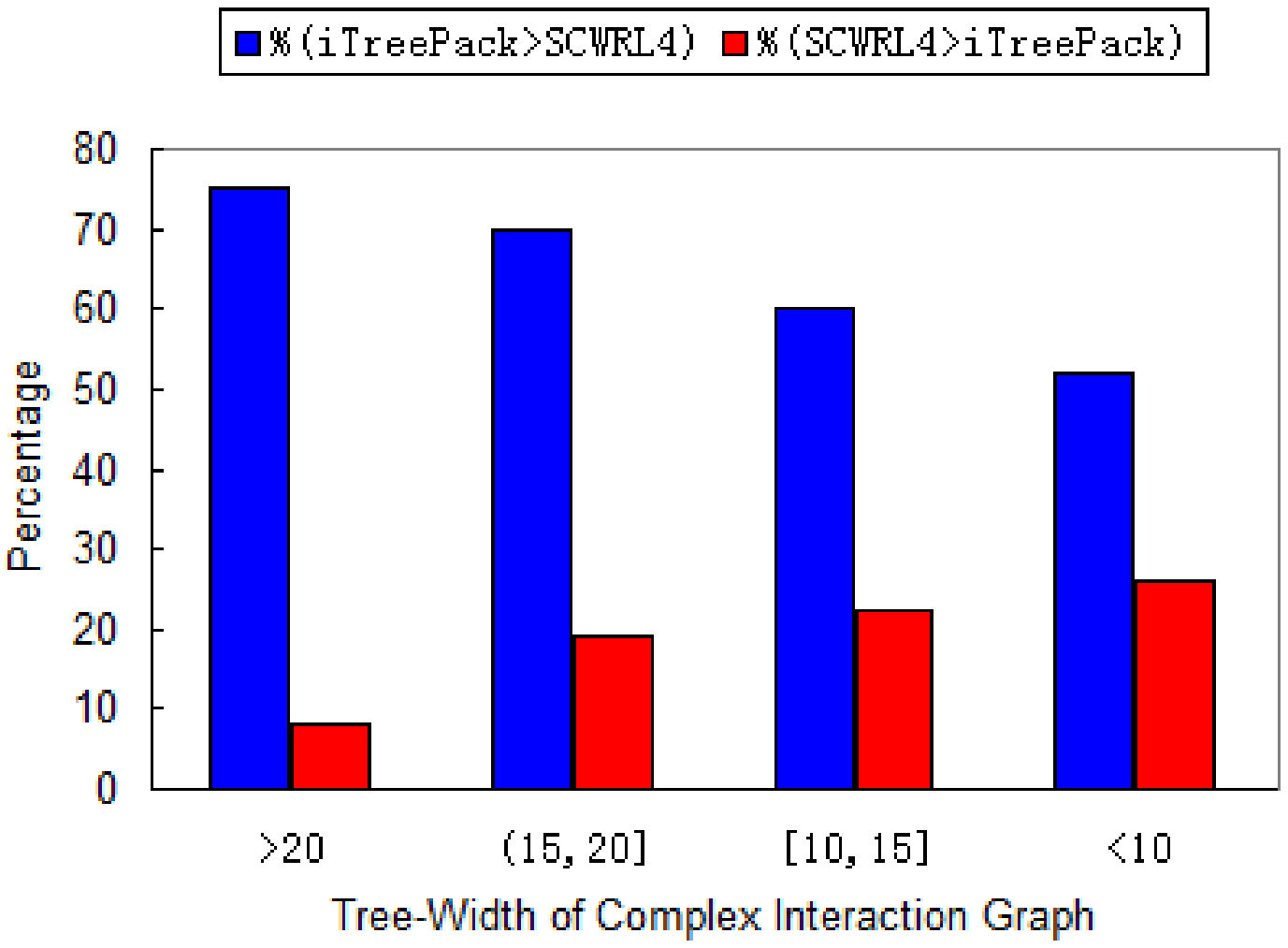} &
\includegraphics[width=0.45\columnwidth]{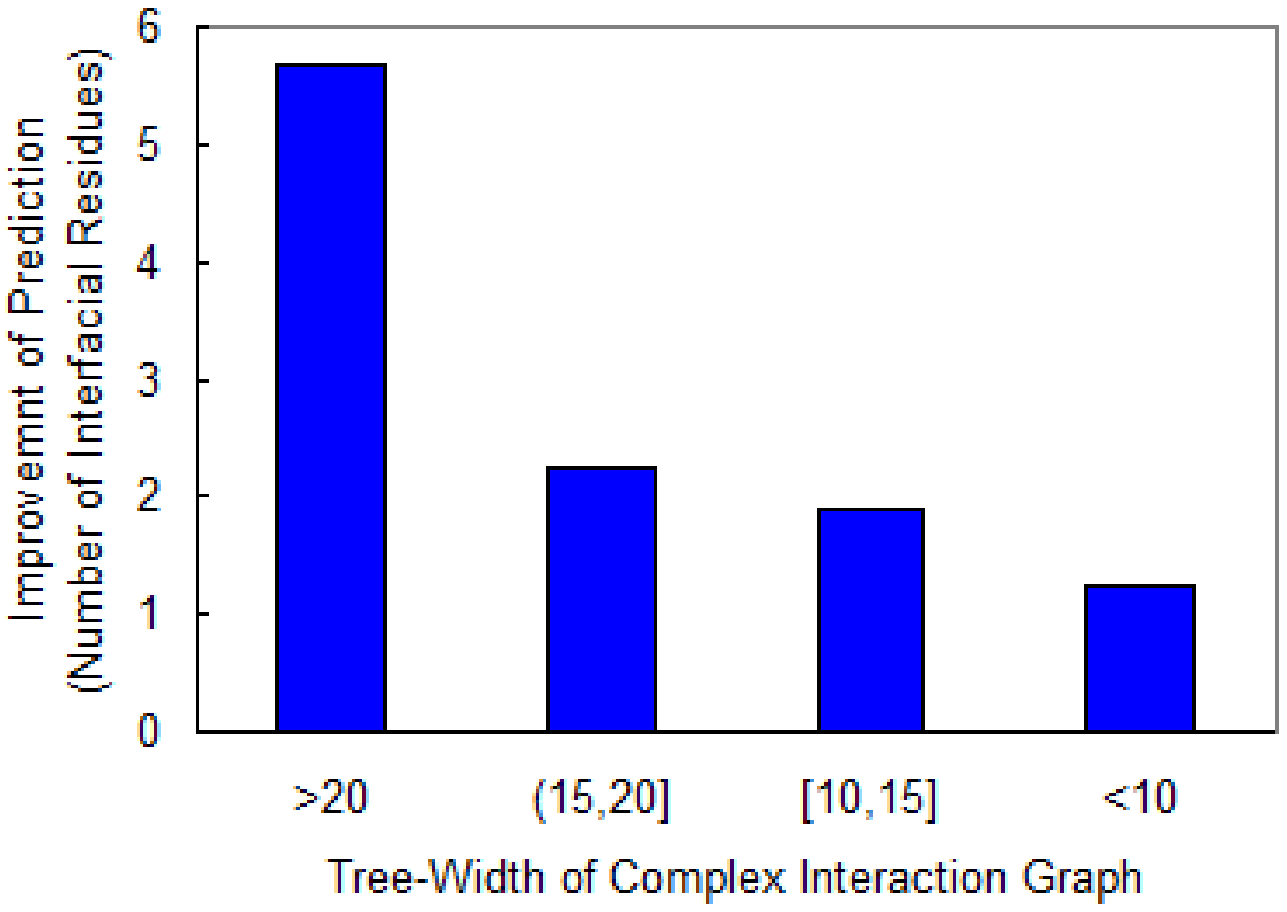} \\
\end{tabular}
\caption{Prediction accuracy of iTreePack and SCWRL4 on interfacial residues.
The y-axis of the left figure shows the percentage of complexes for which one method outperforms the other.
Green bar indicates iTreePack is better while red bar SCWRL4 better.
The right figure shows the margin by which iTreePack is better than SCWRL4,
which is the difference of the average numbers of correctly predicted interfacial residues.}
\label{fg:Comparison}\vspace{-0.2cm}
\end{figure}

\noindent {\bf Prediction Accuracy on Complexes with Large Treewidth.}
Figure \ref{fg:Comparison} shows the accuracy of iTreePack and SCWRL4 on the interfacial residues.
A residue is interfacial if it is in contact with residues in another monomer and
there is a contact between two residues if any of their two respective heavy atoms are within 4.5$\dot{A}$ \cite{PPIPotential}.
As shown in Fig.\ref{fg:Comparison}, iTreePack greatly excels SCWRL4 on the complexes with treewidth larger than 20.
On 122 complexes with treewidth larger than 20, iTreePack outperforms SCWRL4 on about 75\% or 91 of them,
while SCWRL4 does better on only 8\% or 10 of them.
On average, iTreePack correctly predicts side-chain conformations for 5.66 more interfacial residues per complex.
On complexes with treewidth between 16 and 20, iTreePack predicts better side-chain conformations for 70\% of them,
while SCWRL4 does better on 19\%.
On average, iTreePack correctly predicts side-chain conformations for 2.25 more interfacial residues per complex.
Even on those complexes with small treewidth, iTreePack still outperforms SCWRL4 on more than 50\% of them.

\noindent{\bf Case Study.}
iTreePack performs exceptionally well for complexes with very large treewidth.
For example, iTreePack does very well on complex $2oau$, which contains 7 monomers, 21 interfaces, 1442 residues and 543 interfacial residues.
The interaction graph of this complex has treewidth 26, while the maximal treewidth of its constituent monomers is only 10.
SCWRL4-monomer correctly predicts 1063 residues with 379 being interfacial residues.
TreePack correctly predicts 1054  residues with 377 being interfacial.
SCWRL4 correctly predicts 1066 residues with 387 being interfacial, slightly better than SCWRL4-monomer.
In contrast, iTreePack correctly predicts 1083 residues with 428 being interfacial residues.
The physics energy of iTreePack's prediction is $-207kj/mol$ better than SCWRL4's prediction.
That is, there are some serious steric clashes in the side-chain conformations predicted by SCWRL4.

\begin{figure}
\begin{tabular}{ccc}
\includegraphics[width=0.3\columnwidth]{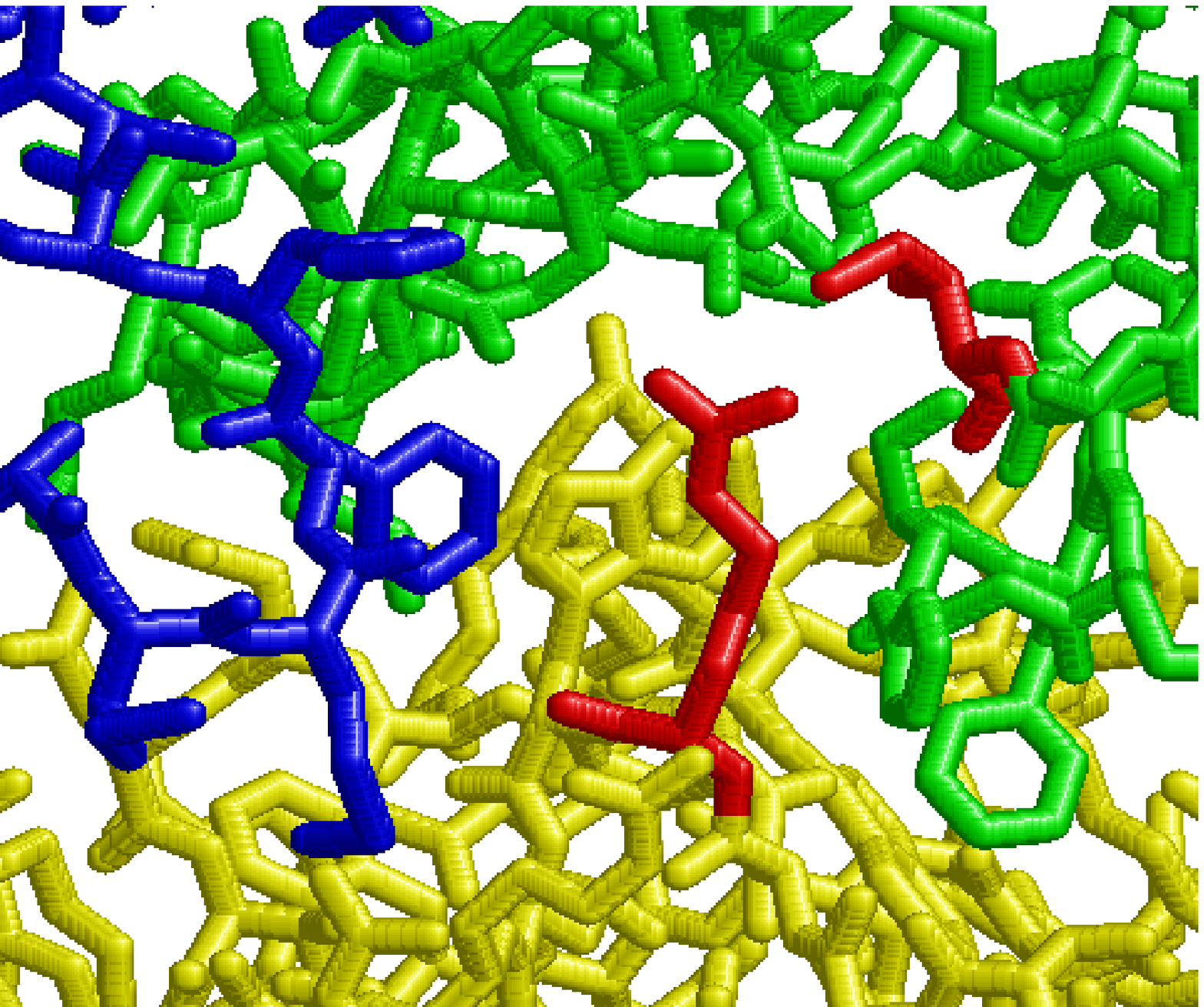} &
\includegraphics[width=0.3\columnwidth]{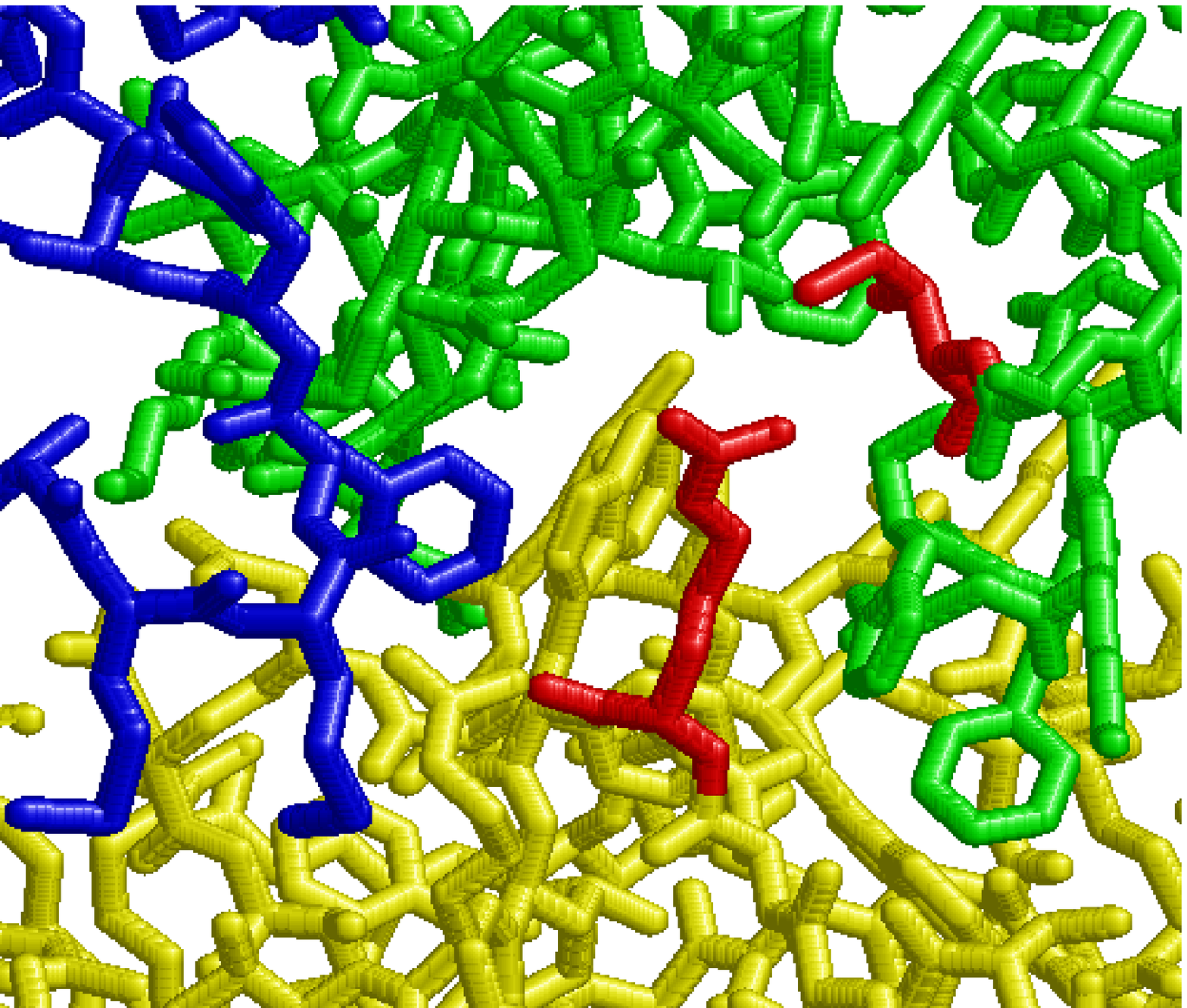} &
\includegraphics[width=0.315\columnwidth]{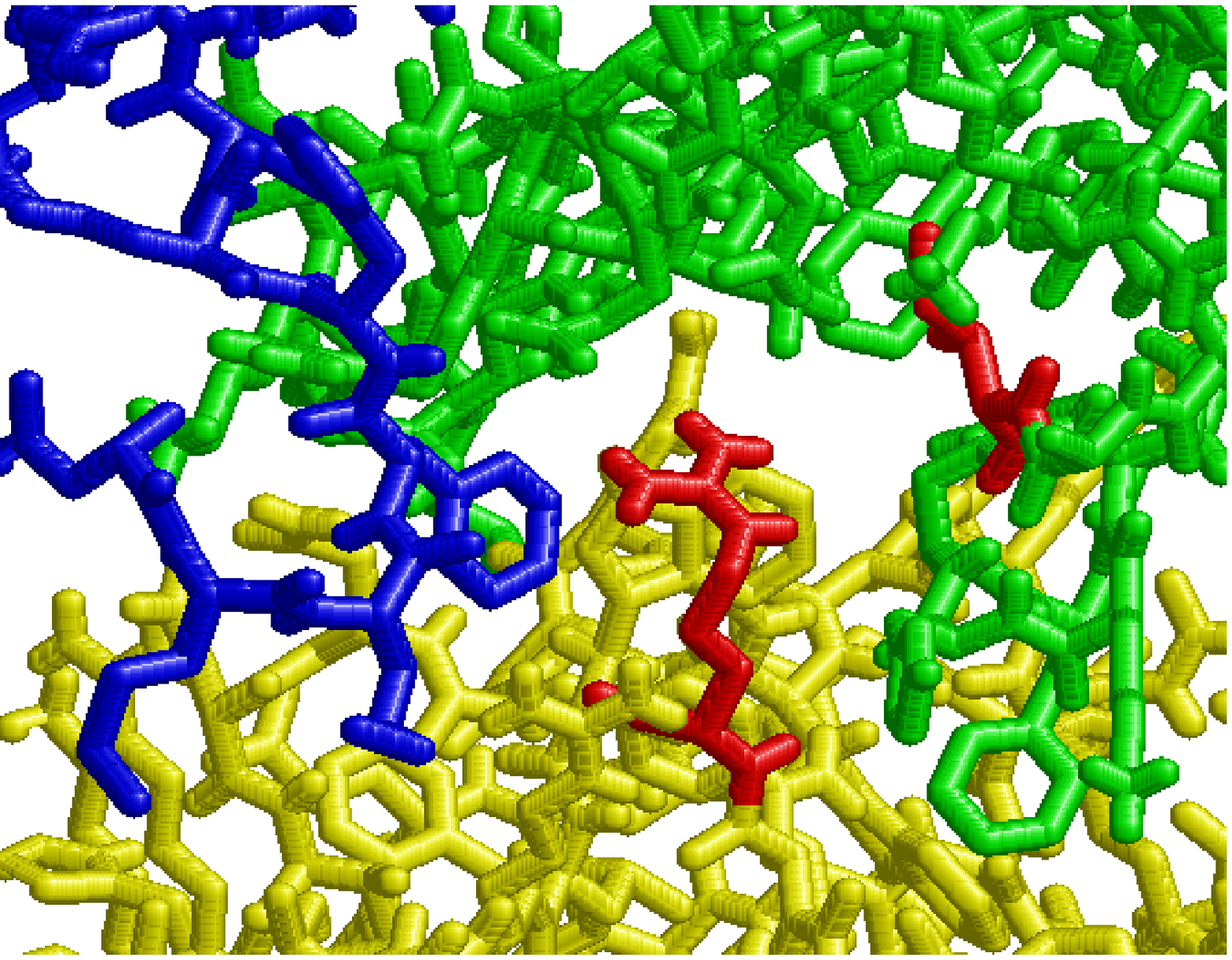} \\
Native Structure & iTreePack & SCWRL4 \\
\end{tabular}
\caption{Side-chain packing on the catalytic interface of adenylyl cyclase and G-protein.
Chains A, B and C are in blue, green and yellow, respectively.
The upper and lower red residues are Arg-986 and Ile-280, respectively.}
\label{fg:Gprt}\vspace{-0.2cm}
\end{figure}

Another example is $1azs$, a complex of G-protein and the catalytic domain of mammalian adenylyl cyclase \cite{GTP}.
In this complex, chains A and B form the adenylate cyclase catalytic domain.
Chain C is a G-protein alpha unit.
These chains closely interact with each other, sharing a common catalytic interface.
The catalytic domain has to bind with alpha unit to perform its enzymic function.
Although three chains are not large, this complex still has large treewidth because of their strong interaction.
The treewidth of the complex is 24 but the maximal treewidth of the three chains is 16.
iTreePack minimizes the energy function to the optimal and correctly predicts 11 more interfacial residues than SCWRL4.
More importantly, several of these 11 residues play very important roles in the catalytic procedure.
Figure \ref{fg:Gprt} shows a pair of interacting residues in the catalytic interface.
In the native structure, Arg-986 of Adenylyl cyclase and Ile-280 of G-protein form a canonical Arg-Ile interaction.
Both residues are also spatially close to many other residues in all three chains.
iTreePack predicts their side-chain conformations almost perfectly and fully recovers the important interaction.
However, SCWRL4 fails to predict the side-chain conformations of these residues and results in several steric clashes.
{
\begin{table*}[ht]
\caption{Prediction accuracy on the 547 complexes and the SCWRL4 dataset of 379 monomers.}
\begin{center}
\begin{tabular}{|l|c|c|c|c|}
\hline
Methods   & All Residues & Non-interfacial Residues & Interfacial Residues  & SCWRL4 dataset \\
\hline
SCWRL4-monomer & 78.07\% & 78.31\% & 76.41\% &  85.22\% \\
\hline
TreePack & 78.20\% & 78.44\% & 76.81\% &  85.87\% \\
\hline
SCWRL4 & 80.28\% & 79.61\% & 82.55\% & 85.22\% \\
\hline
iTreePack & {\bf 81.98\%} & {\bf 81.11\%} & \bf{84.91\%} & \bf{85.87\%} \\
\hline
\end{tabular}
\label{table:result}
\end{center}\vspace{-0.2cm}
\end{table*}
}

\noindent{\bf Overall Prediction Accuracy.}
The 4th column of Table \ref{table:result} shows the average prediction accuracy on the interfacial residues.
TreePack and SCWRL4-monomer perform poorly since they do not take into consideration the interactions among monomers.
In contrast, iTreePack and SCWRL4 produce substantially better predictions (6-8\% better).
On the interfacial residues, iTreePack is about 2.36\% better than SCWRL4.
On the non-interfacial residues, iTreePack is 1.5\% better than SCWRL4.
This indicates that iTreePack's dual decomposition technique can optimize the energy function more accurately than SCWRL4 and thus, results in better side-chain packing.

\begin{table}
\caption{Average running time of iTreePack and SCWRL4 on the 547 complexes and the SCWRL4 dataset of 379 monomers.}
\begin{center}
\begin{tabular}{|l|c|c|}
\hline
Methods   & Complex Dataset  & SCWRL4 dataset \\
\hline
SCWRL4    &  32s & 7.6s\\
\hline
iTreePack & 44s & 1.8s \\
\hline
\end{tabular}
\label{table:result-time}
\end{center}\vspace{-0.7cm}
\end{table}
Note that in terms of average prediction accuracy iTreePack is not significantly better than SCWRL4 because many test complexes
have small treewidth and iTreePack has no advantage over SCWRL4 on these complexes.
However, as shown in previous sections, iTreePack is much better than SCWRL4 on complexes with large treewidth.

The third column of Table \ref{table:result} shows the accuracy on all the residues of the 547 complexes.
TreePack and SCWRL4-monomer perform similarly and worse (about 2-3\%) than iTreePack and SCWRL4.
If only non-interfacial residues are considered, TreePack and SCWRL4-monomer are still worse than iTreePack and SCWRL4.
This implies that it is important to consider inter-monomer interactions when predicting side-chain conformations for protein complexes.
iTreePack is 1.7\% better than SCWRL4 on all the residues, while TreePack is only 0.13\% than SCWRL4-monomer.
On average, van der Waals' energy of the side-chain placement by iTreePack is lower than that of SCWRL4 by $9.98kcal/mol$.
iTreePack produces more energetically favorable side-chain packing than SCWRL4 for $96.8\%$ complexes.

Finally, tested on the SCWRL4 dataset of 379 monomer proteins, iTreePack/TreePack is marginally better than SCWRL4 on $\chi_1$ accuracy.


\noindent{\bf Running Time.}
We tested iTreePack, SCRWL4 and TreePack using all the 547 complexes on a single core of Intel 2.4GHZ Xeon CPU with 8G RAM.
TreePack fails to finish side-chain packing for each of 81 complexes within 1 hour.
For the remaining complexes, the average running time of TreePack is 17.8 minutes.
By contrast, iTreePack can solve the side-chain packing problem for all the 547 complexes,
with the average running time being only 44 seconds,
as shown in Table \ref{table:result-time}.
SCWRL4 has an average running time of 32 seconds, slightly faster than iTreePack, but at the cost of packing accuracy.

Tested on the 379 monomer proteins of the SCWRL4 dataset, iTreePack is four times faster than SCWRL4 and slightly more accurate.
In particular, iTreePack has an overall running time of 675 seconds while SCWRL4 2897 seconds.
\section{Conclusion}
We have presented an efficient algorithm, iTreePack, for protein complex side-chain packing.
This method first converts the complex side-chain packing problem into a dual relaxation problem,
which can be decomposed into a few small subproblems.
Each corresponds to the side-chain packing of  a monomer or a protein interface
and can be solved by tree-decomposition efficiently and separately.
A projected subgradient descent algorithm is then applied to assembling the solutions of the subproblems
into a single coherent solution of the original problem.
This algorithm can solve the complex side-chain packing problem much more efficiently and accurately
than the pure tree-decomposition algorithm and thus, can do side-chain packing for very large complexes.
Experimental results show that iTreePack can place side-chain atoms much more accurately than SCWRL4 for very large complexes
since iTreePack can optimize the energy function much better than SCRWL4.
Such a dual decomposition algorithm can also be applied to speeding up side-chain packing of large monomer proteins,
which usually contain several domains.
That is, we can first do side-chain packing for each domain separately
and then assemble the side-chain packings of all the domains into a packing of the whole monomer protein.

The experimental results also show that there is still a gap between the side-chain packing accuracy of a complex and monomer protein.
This may be because that we use the same rotamer library for both non-interfacial and interfacial residues and a simple energy function
adapted from monomer side-chain packing.
In the future,
we are going to develop a rotamer library for interfacial residues and better energy functions for complex side-chain packing.\\

\noindent{\bf Acknowledgments.}
This work is supported by National Institutes of Health grants R01GM0897532 and R01GM081871,
National Science Foundation DBI-0960390.

\bibliographystyle{plain}
\bibliography{SideChain}

\end{document}